\documentclass[aps,prc,twocolumn,showpacs,nofootinbib]{revtex4-1}
\usepackage[utf8]{inputenc}
\usepackage{graphicx}   
\usepackage{latexsym}   
\usepackage{enumerate}
\usepackage{rotating,booktabs,multirow}
\usepackage{amsmath}
\usepackage{amsfonts}
\usepackage{amssymb}
\usepackage{multirow}
\usepackage{bm}
\usepackage{xcolor}

\begin{document}
\title{Harmonic flow correlations in Au+Au reactions at 1.23 AGeV:\\ A new testing ground for the Equation-of-State and expansion geometry}
\author{Tom Reichert$^{1,3}$, Jan~Steinheimer$^{5}$, Christoph Herold$^4$, Ayut Limphirat$^4$, Marcus~Bleicher$^{1,2,3}$}

\affiliation{$^1$ Institut f\"ur Theoretische Physik, Goethe Universit\"at Frankfurt, Max-von-Laue-Strasse 1, D-60438 Frankfurt am Main, Germany}
\affiliation{$^2$ GSI Helmholtzzentrum f\"ur Schwerionenforschung GmbH, Planckstr. 1, 64291 Darmstadt, Germany}
\affiliation{$^3$ Helmholtz Research Academy Hesse for FAIR (HFHF), GSI Helmholtz Center for Heavy Ion Physics, Campus Frankfurt, Max-von-Laue-Str. 12, 60438 Frankfurt, Germany}
\affiliation{$^4$ Suranaree University of Technology, University Avenue 111, Nakhon Ratchasima 30000, Thailand}
\affiliation{$^5$ Frankfurt Institute for Advanced Studies (FIAS), Ruth-Moufang-Str.1, D-60438 Frankfurt am Main, Germany}

\begin{abstract}
Correlations between the harmonic flow coefficients $v_1$, $v_2$, $v_3$ and $v_4$ of nucleons in semi-peripheral Au+Au collisions at a beam energy of 1.23~AGeV are investigated within the hadronic transport approach Ultra-relativistic Quantum Molecular Dynamics (UrQMD). In contrast to ultra-relativistic collision energies (where the flow coefficients are evaluated with respect to the respective event plane), we predict strong correlations between the flow harmonics with respect to the reaction plane. Based on an event-by-event selection of the midrapidity final state elliptic flow of nucleons we show that as a function of rapidity, I) the sign of the triangular flow changes, II) that the shape of $v_4$ changes from convex to concave, and III) that $v_3\propto v_1v_2$ and $v_4\propto v_2^2$ for all different event classes, indicating strong correlations between all investigated harmonic flow coefficients.  
\end{abstract}

\maketitle

\section{Introduction}
The exploration of the properties of dense and hot nuclear matter is an ongoing endeavor since more than 40 years. Nowadays, the main focus lies in two areas: I) The systematic investigation of the phase diagram of Quantum-Chromo-Dynamics to understand the onset of deconfinement \cite{Gazdzicki:2010iv} and II) the extraction of the nuclear matter Equation-of-State (EoS) at moderate temperatures and its relation to astrophysical objects \cite{Durante:2019hzd}. On the experimental side, the Large Hadron Collider (LHC) probes the strong interaction at the highest available collision energies. Here, a deconfined and nearly net-baryon density free system at very high temperatures is created, and such an environment is theoretically well accessible by calculations of Quantum-Chromo-Dynamics (QCD) on the lattice. At this high-temperature frontier the Fourier decomposition of the azimuthal angle distribution of emitted particles can shed light on the expansion dynamics of the created matter. Especially, the elliptic flow (usually called $v_2$) is a fabric of pressure gradients in the transverse direction and allows to extract the viscosity (and other transport coefficients) of the Quark-Gluon-Plasma (QGP). The studies at the LHC extend and complement previous and ongoing measurements at Relativistic Heavy Ion Collider (RHIC) \cite{STAR:2000ekf,PHENIX:2003iij}, which have pioneered the extraction of the viscosity of a Quark-Gluon-Plasma (QGP) and have lead to the strongly coupled perfect liquid hypothesis of the QGP \cite{Huovinen:2001cy,Song:2007fn,Romatschke:2007mq,Luzum:2008cw}. Further studies at RHIC and at LHC on the triangular ($v_3$) and quadrangular ($v_4$) flow  have tied both coefficients to the initial state fluctuations \cite{ALICE:2011ab,Alver:2010gr,Petersen:2010cw} which can provide further insights on the parton structure of the impinging nuclei \cite{Schenke:2010rr}. 

With decreasing energy, the high-density frontier is probed. Here prominent examples are the RHIC beam energy scan program (RHIC-BES) \cite{Ma:2022jxx} and the future FAIR \cite{Klochkov:2021eyo} and NICA \cite{MPD:2022qhn} facilities. In this energy domain, the onset of deconfinement is expected and a gradual transition (with increasing beam energy) to a QGP is expected. At even lower beam energies, e.g. currently actively explored by the GSI facility, the nuclear Equation-of-State (EoS) of hadronic matter \cite{Reisdorf:2002hw} is explored and will allow to bridge the gap to binary neutron star mergers, which might provide complementary information on the EoS via the detection of gravitational waves \cite{LIGOScientific:2018cki}. In this energy regime, the Fourier decomposition of the azimuthal angle distribution of the created particles is driven by an intricate interplay between the EoS \cite{Danielewicz:2002pu,Hillmann:2018nmd}, time dependent expansion dynamics/geometry \cite{Sorge:1998mk} and viscous corrections \cite{Reichert:2020oes}. Precise measurements up to the sixth flow coefficient of protons and light clusters ($d$, $t$) have already been successfully accomplished by the HADES collaboration at GSI \cite{HADES:2020lob}. 

A novel tool to explore the properties and geometry of the created QCD matter, namely harmonic flow correlations, was recently suggested and measured in ultra-relativistic nucleus-nucleus reactions at RHIC and at LHC \cite{ALICE:2016kpq,STAR:2018fpo}. At both energies an anti-correlation between event-by-event fluctuations of $v_2$ and $v_3$ was observed, while the event-by-event fluctuations of $v_2$ and $v_4$ were found to be correlated. These results were also confirmed by hydrodynamic \cite{Niemi:2012aj} and transport simulations \cite{Zhou:2015eya,Zhu:2016puf}. Generally, the (anti-)correlations were traced-back to the initial state eccentricities and it was concluded that the investigation of the correlations between different flow harmonics have a substantially higher sensitivity to the details of theoretical calculations (transport coefficients, equation-of-state, initial state modeling) in comparison to individual $v_n$ coefficients.

In this article we explore for the first time the event-by-event correlations between the first four harmonic flow coefficients ($v_1$ to $v_4$) to extract further information on the properties and geometries of the matter created at low beam energies. We propose to test these predictions with the currently running HADES experiment.

\section{Model setup and flow extraction}
For the present study we use the Ultra-relativistic Quantum Molecular Dynamics (UrQMD) model \cite{Bass:1998ca,Bleicher:1999xi,Bleicher:2022kcu} in its most recent version (v3.5). UrQMD is a dynamical microscopic transport simulation based on the explicit propagation of hadrons in phase-space. The imaginary part of the interactions is modeled via binary elastic and inelastic collisions, leading to resonance excitations and decays or color flux-tube formation and their fragmentation. The real part of the interaction potential is implemented via different equations of state (following the usual notion of a hard and soft equation-of-state), alternative equations-of-state, e.g. a chiral mean field EoS can also be introduced, see e.g. \cite{Kuttan:2022zno}. In its current version, UrQMD includes a large body of baryonic and mesonic resonances up to masses of 4~GeV. The model is well established in the GSI energy regime. For recent studies of the bulk dynamics, we refer the reader to \cite{Reichert:2020uxs,Reichert:2021ljd}. For the analysis of the integrated harmonic flows at GSI energies see \cite{Hillmann:2018nmd,Hillmann:2019wlt}. 

The flow coefficients are identified with the Fourier coefficients in the series expansion of the azimuthal angle distribution which can be written as
\begin{equation}
\begin{aligned}
    \frac{\mathrm{d}N}{\mathrm{d}\phi} = 1 + 2\sum\limits_{n=1}^\infty v_n\cos(n(\phi-\Psi_{RP}))& \\ 
    + \tilde{v}_n\sin(n(\phi-\Psi_{RP}))& ,
\end{aligned}
\end{equation}
in which $v_n$ is n-th order flow coefficient of the even (cosine) term, $\tilde{v}_n$ is the n-th order flow coefficient of the odd (sine) term, $\phi$ is the azimuthal angle and $\Psi_{RP}$ is the angle of the reaction plane. The HADES experiment uses a forward wall to reconstruct the event plane from the spectator nucleons \cite{Kardan:2017knj,HADES:2020lob}. In the simulation, the reaction plane is fixed because the impact parameter is known and thus $\Psi_{RP}=0$ is used for the present analysis of the simulation. However, the first order spectator event plane can still fluctuate on an event by event basis which is reflected in general in nonzero sine terms for a single event. But by taking the average over all events (in a given event class) they will be zero due to symmetry. We hence restrict our investigation to the cosine terms (i.e. the projection of the flow vector of each particle onto the known reaction plane). The flow coefficients are then calculated as
\begin{equation}
    v_n = \langle \cos(n(\phi-\Psi_{RP})) \rangle,
\end{equation}
where the average $\langle\cdot\rangle$ is taken over all nucleons in a fixed rapidity or transverse momentum range in a given event.


Let us stress that the extraction of the flow coefficents in the HADES experiment is different from the methods used at higher energies. At high energies, e.g. at RHIC and LHC, the flow coefficients are usually extracted with respect to the n-th order event plane which is constructed via the flow vector $Q$ \cite{Poskanzer:1998yz,Borghini:2000sa}. Another possibility is to extract the flow from the 2- and 4-particle cumulants \cite{Borghini:2001vi} or to use the Lee-Yang zeroes method \cite{Bhalerao:2003xf}. Generally such methods allow to obtain the magnitude of the flow coefficients and work best (especially in case of the Lee-Yang zero method), if the multiplicites are sufficiently high and the flow harmonics are not too small \cite{Zhu:2006fb}. At low energies the situation is different, the multiplicities are limited, the flow coefficients might be rather small and the sign of the elliptic flow is important (because of the change from in-plane to out-of-plane  emission (squeeze-out) due to the blocking by the spectators towards lower energies). Therefore, the HADES experiment uses a different method and extracts the coefficients with respect to the reaction plane. In the simulation this can be done exactly, because the reaction plane is known. However in the experiment the reaction plane has to be reconstructed. The HADES collaboration uses the first order spectator event plane as a good estimator for the reaction plane with high resolution \cite{Kardan:2017knj,HADES:2020lob}.

Previously, the flow correlations have been quantified using Pearson correlation functions \cite{Niemi:2012aj}. However, for the present study we focus mainly on different event classes straightforwardly and show the correlation directly, because showing the full distributions allows for a more direct interpretation of the results than compressing the distribution into a single number. However, at the end of the paper, we also provide the Pearson coefficients to allow for a comparison to the data at higher energies.

\section{Results}
All results were obtained by simulating 20-30\% peripheral Au+Au collisions at E$_{\rm beam}=1.23$~AGeV kinetic beam energy with the UrQMD (v3.5) model. The centrality is selected via impact parameter cuts following previous Monte-Carlo Glauber simulations \cite{HADES:2017def}. We employ the model mainly with a hard equation-of-state as it was shown to yield the best description of the measured HADES data (cf. Ref. \cite{Hillmann:2018nmd,Hillmann:2019wlt}). We focus our analysis on participating nucleons and also exclude nucleons that are bound in light clusters. It has been shown \cite{Hillmann:2019wlt} that both effects need to be taken into account to reliably describe the measured data \cite{HADES:2020lob}.

\subsection{Elliptic flow fluctuations and event class selection}
\begin{figure}
    \centering
    \includegraphics[width=\columnwidth]{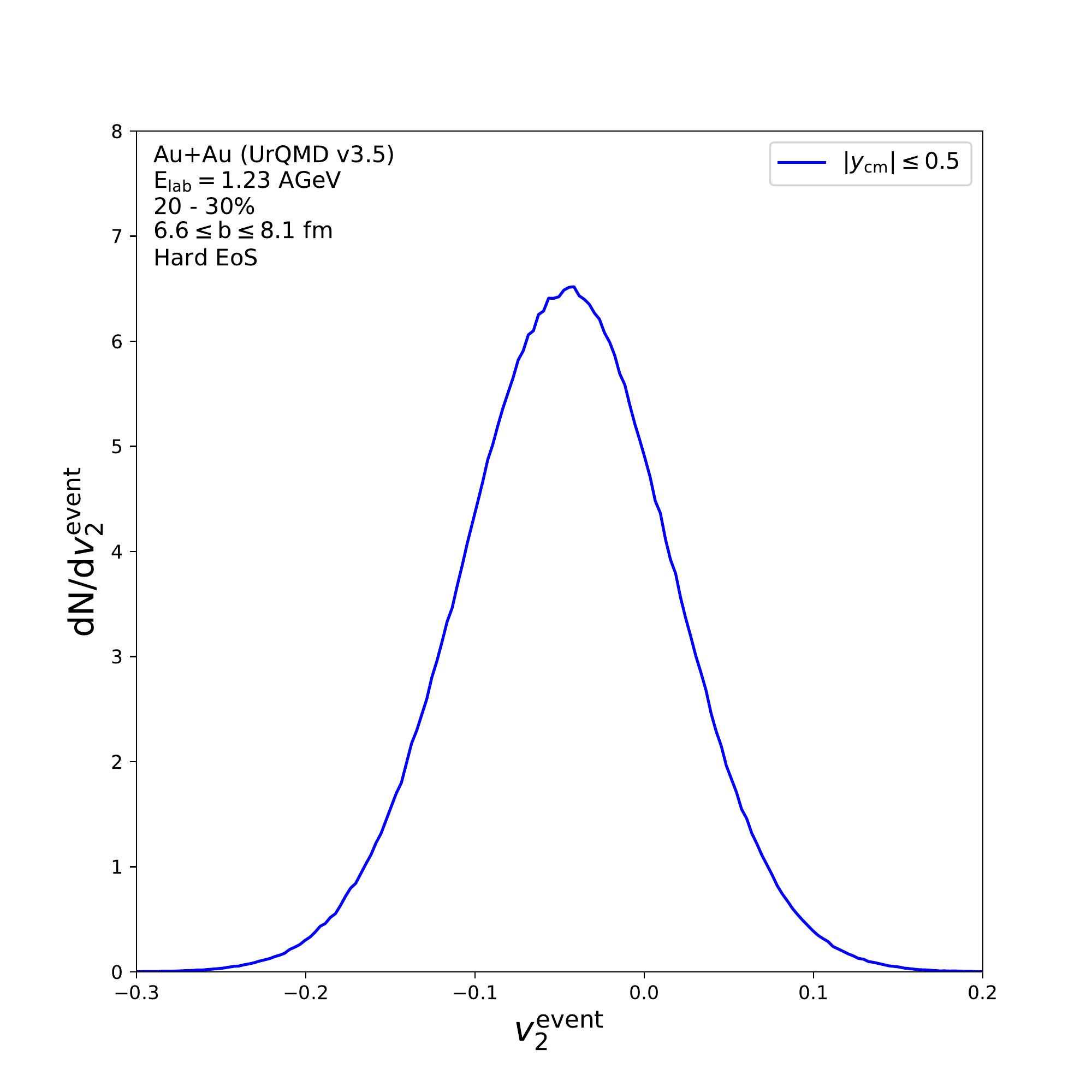}
    \caption{[Color online] The $v_2$ distribution integrated over all nucleons in the rapidity window $|y_\mathrm{cm}|\leq0.5$ in 20-30\% central Au+Au collisions at kinetic beam energy of 1.23~AGeV from UrQMD with a hard EoS.}
    \label{fig:dNdv2_event}
\end{figure}

We start our investigation with an analysis of the event-by-event distribution of the final state elliptic flow. Fig. \ref{fig:dNdv2_event} shows the $v_2$ distribution integrated over all nucleons in the rapidity window\footnote{In this paper we work in the center-of-mass frame where midrapidity is at $y_{cm}=0$.} $|y_{cm}|\leq0.5$ in 20-30\% central Au+Au collisions at kinetic beam energy of 1.23~AGeV from UrQMD with a hard EoS. We observe that the event-by-event value of the elliptic flow fluctuates substantially around the mean value of $\langle v_2\rangle_{|y_\mathrm{cm}|\leq0.5}$ = -0.05 (the mean value is in line with the HADES data \cite{HADES:2020lob} and also consistent with older data in a similar energy region \cite{FOPI:2004bfz,E895:1999ldn,E895:2000maf}). The width of the $v_2$ distribution is rather broad and has a FWHM of 0.15. Thus, single events can even show an overall positive $v_2$, as well as highly negative $v_2$ values. This finding naturally gives rise to the idea that heavy-ion collision events can be categorized into different classes based on their final elliptic flow around midrapidity, while keeping everything else fixed. Such a categorization is well known from ultra-relativistic collisions under the name ``event shape engineering'' \cite{Schukraft:2012ah} connecting event wise final flow to the initial geometric (spatial) configuration. Although the correlation between initial spatial fluctuations and final elliptic flow is less obvious at low energies than at ultra-relativistic energies, due to the intricate emission dynamics between spectator blocking and expansion, the observation of strong final state elliptic flow fluctuations opens new possibilities for quantitative tests of the properties of high density QCD matter and the dynamical models used for its investigation. 

\subsection{Flow in different event classes}
Let us now select event classes based on the event-wise elliptic flow at midrapidity, but keeping everything else (collision system, energy and centrality) fixed. To this aim, we split the events into 14 classes defined by the event wise integrated elliptic flow at midrapidity, $\langle v_2\rangle_{|y_\mathrm{cm}|\leq0.5}$ with the specific values for the (momentum-space-)ellipticity given in Tab. \ref{table:eventclasses}.
\begin{table}[!ht]
\centering
\renewcommand{\arraystretch}{1.2} 
\begin{tabular}{|c|c|}
\hline
event class & ellipticity \\
\hline
\hline
1 & $\langle v_2\rangle_{|y_\mathrm{cm}|\leq0.5} \leq -0.200$ \\
\hline
2 & $-0.200 \leq \langle v_2\rangle_{|y_\mathrm{cm}|\leq0.5} \leq -0.175$ \\
\hline
3 & $-0.175 \leq \langle v_2\rangle_{|y_\mathrm{cm}|\leq0.5} \leq -0.150$ \\
\hline
4 & $-0.150 \leq \langle v_2\rangle_{|y_\mathrm{cm}|\leq0.5} \leq -0.125$ \\
\hline
5 & $-0.125 \leq \langle v_2\rangle_{|y_\mathrm{cm}|\leq0.5} \leq -0.100$ \\
\hline
6 & $-0.100 \leq \langle v_2\rangle_{|y_\mathrm{cm}|\leq0.5} \leq -0.075$ \\
\hline
7 & $-0.075 \leq \langle v_2\rangle_{|y_\mathrm{cm}|\leq0.5} \leq -0.050$ \\
\hline
8 & $-0.050 \leq \langle v_2\rangle_{|y_\mathrm{cm}|\leq0.5} \leq -0.025$ \\
\hline
9 & $-0.025 \leq \langle v_2\rangle_{|y_\mathrm{cm}|\leq0.5} \leq +0.000$ \\
\hline
10 & $+0.000 \leq \langle v_2\rangle_{|y_\mathrm{cm}|\leq0.5} \leq +0.025$ \\
\hline
11 & $+0.025 \leq \langle v_2\rangle_{|y_\mathrm{cm}|\leq0.5} \leq +0.050$ \\
\hline
12 & $+0.050 \leq \langle v_2\rangle_{|y_\mathrm{cm}|\leq0.5} \leq +0.075$ \\
\hline
13 & $+0.075 \leq \langle v_2\rangle_{|y_\mathrm{cm}|\leq0.5} \leq +0.100$ \\
\hline
14 & $+0.100 \leq \langle v_2\rangle_{|y_\mathrm{cm}|\leq0.5} $\\
\hline
\end{tabular}
    \caption{Definition of the event classes based on the elliptic flow value (ellipticity) at midrapidity in semi-peripheral Au+Au collisions at 1.23~AGeV using UrQMD with a hard EoS.}
    \label{table:eventclasses}
\end{table}

\begin{figure}
    \centering
    \includegraphics[width=\columnwidth]{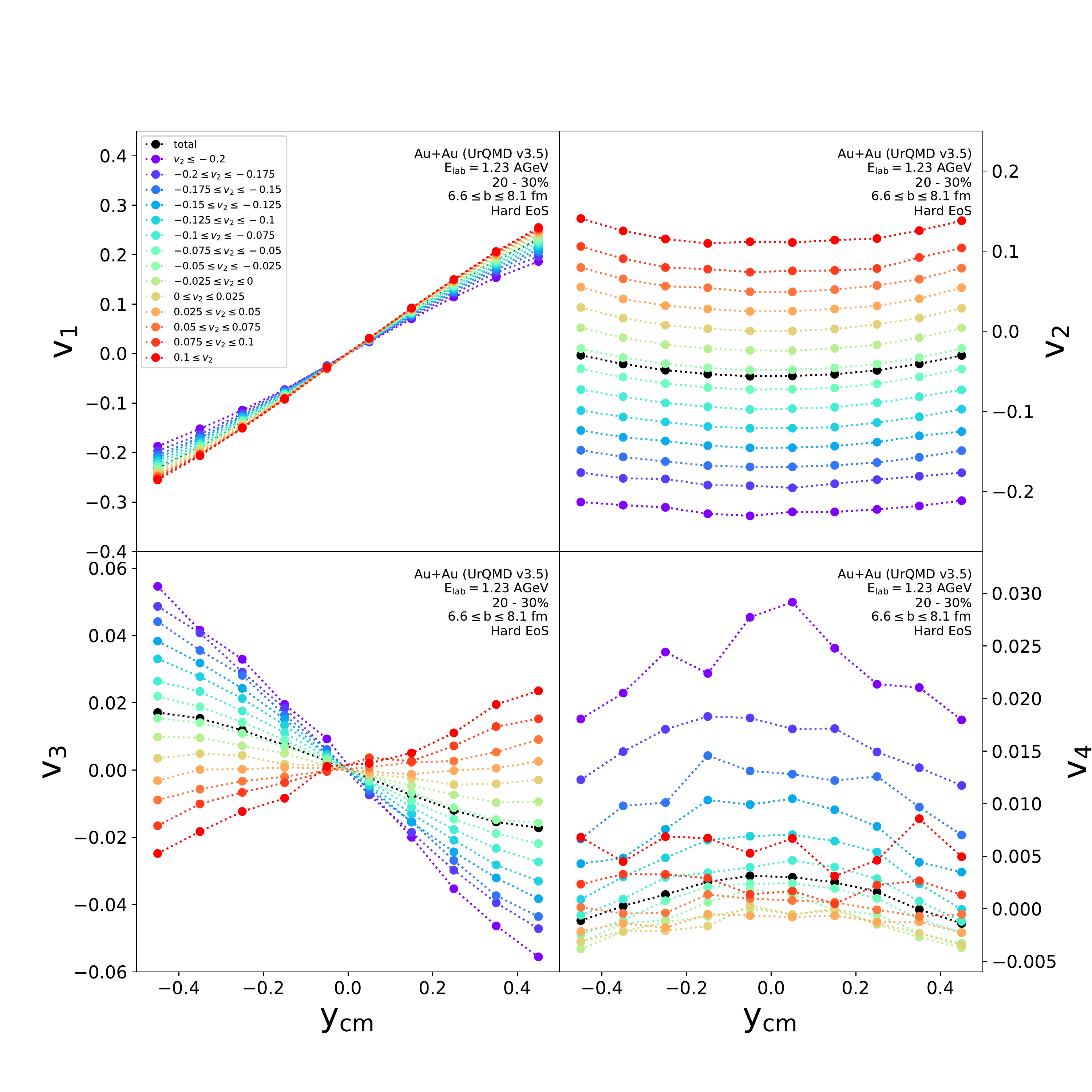}
    \caption{[Color online] Rapidity dependence of the flow coefficients $v_1$ (upper left), $v_2$ (upper right), $v_3$ (lower left) and $v_4$ (lower right) in different event classes (from the most positive $v_2$ class in red to most negative $v_2$ class in purple, see legend) as well as the total (i.e. without event class selection) flow coefficient (black) for 20-30\% peripheral Au+Au collisions with hard EoS at 1.23~AGeV from UrQMD.}
    \label{fig:v1v2v3v4_event_class}
\end{figure}

For each of these event classes, we now investigate the first to fourth order flow coefficients extracted in each specified event class as shown in Fig. \ref{fig:v1v2v3v4_event_class} for semi-peripheral Au+Au collisions at 1.23~AGeV using a hard EoS. The top left figure shows the rapidity dependence of the flow coefficient $v_1$, the top right figure shows $v_2$, the bottom left figure shows $v_3$ and the bottom right figure depicts $v_4$. The event shape classes are denoted by colors from the most negative $v_2$ event class (event class 1) in purple to the most positive $v_2$ event class (event class 14) in red as well as the unbiased distributions in black. 

\begin{figure}
    \centering
    \includegraphics[width=\columnwidth]{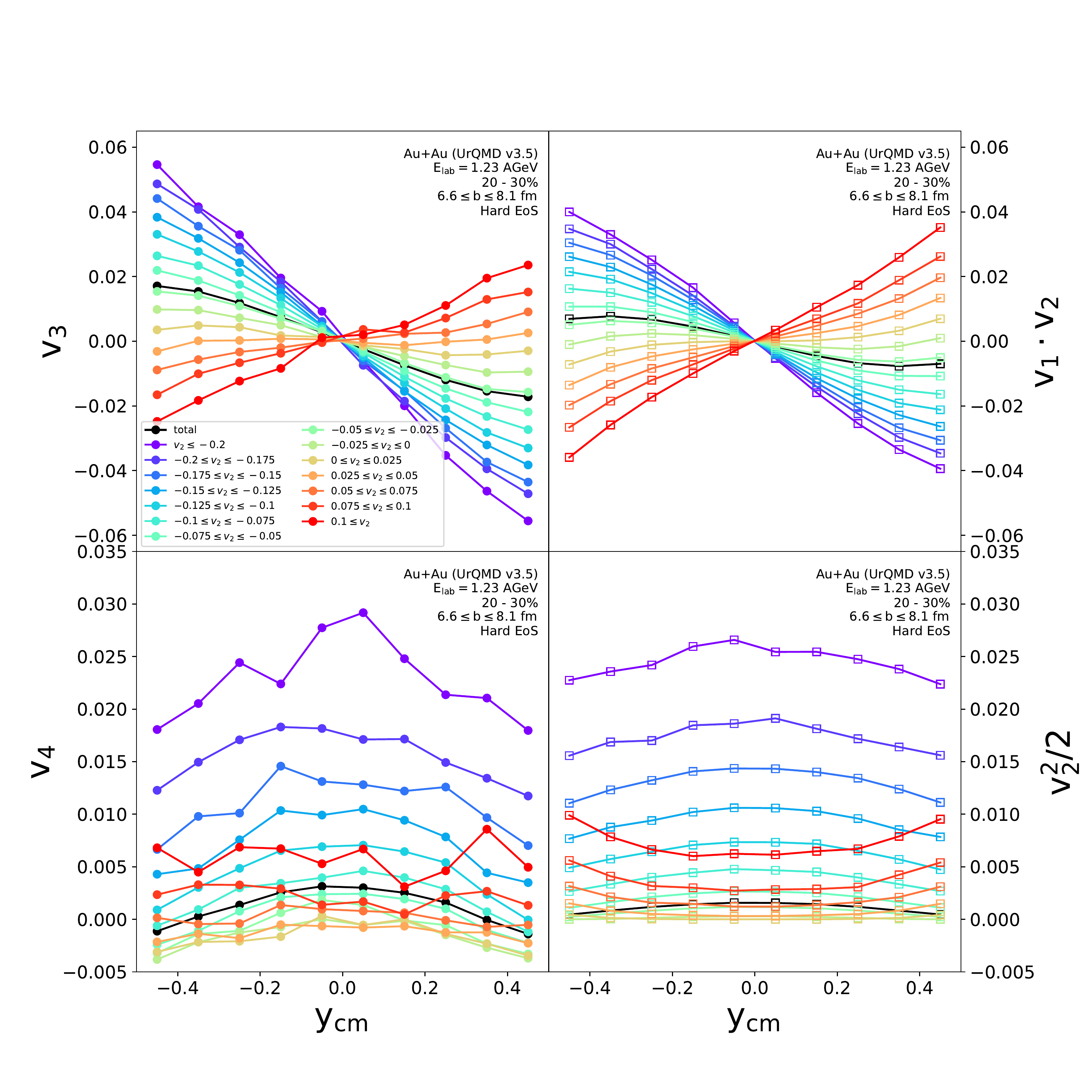}
    \caption{[Color online] The rapidity dependence of triangular flow $v_3$ (upper left) in comparison to the multiplicative relation $v_1\cdot v_2$ (upper right) and the rapidity dependence of quadrangular flow $v_4$ (lower left) in comparison to the scaling relation $0.5v_2^2$ (lower right) in different event classes (from the most positive $v_2$ class in red to most negative $v_2$ class in purple, see legend) as well as the total (i.e. without event class selection) flow coefficient (black) from 20-30\% peripheral Au+Au collisions with hard EoS at 1.23~AGeV from UrQMD.}
    \label{fig:scaling_relations}
\end{figure}

Let us begin the discussion with the directed flow $v_1$. We observe that the directed flow is only marginally affected by the selection of the event class. This is expected, because the directed flow is mainly driven by the bounce-off of the impinging nuclei. Nevertheless, a small dependence on the $v_2$ event class can be observed with a positive correlation leading to a larger magnitude of $v_1$ at forward/backward rapidities (i.e. an increase of ${\rm d} v_1/{\rm d}y_\mathrm{cm}|_{y_\mathrm{cm}=0}$) with the selection of higher $v_2$ values. Turning to the elliptic flow, we observe that the event class selection via the final state $v_2$ shifts the magnitude of the elliptic flow, while its qualitative rapidity dependence stays unaffected. The major effects are however found when exploring the triangular flow and its correlation to $v_2$. The magnitude and rapidity dependence of $v_3$ vary drastically with different $v_2$ event classes. While for averaged (i.e. unselected) events one finds a slightly negative $v_2$ and also a $v_3$ with a slightly negative slope at midrapidity, for positive elliptic flow event classes the triangular flow develops a strongly positive slope. Also for the opposite event class (strongly negative $v_2$) we observe a correlation indicated by a steeper negative slope of $v_3$ for such events. This clearly shows the strong correlation between $v_2$ and $v_3$ in the HADES energy regime. Finally, we address the rapidity dependence of the quadrangular flow coefficient $v_4$. Also $v_4$ shows an interesting and unexpected behavior: While events in the negative $v_2$ event class show a concave shape of $v_4$ as function of rapidity, event classes with a positive $v_2$ tend to develop a convex shape in rapidity. This indicates that also $v_2$ and $v_4$ are intertwined even on an event-by-event basis and show a strong anti-correlation (larger $v_2$ leads to smaller $v_4$ at midrapidity). One should note that the correlations among the flow coefficients are strikingly different at low energies in comparison to high energies and also much stronger pronounced at the low energies investigated here. The flow correlations found here are attributed to the huge influence of the expansion geometry and the intricate space and time dependent emission (blocking of the expansion in impact parameter direction during the overlap phase leading to negative $v_2$, followed by an expansion in impact parameter direction resulting in a positive $v_2$) pattern which are fundamentally different than at high energies where the expansion in the impact parameter direction dominates already from the beginning.

\subsection{Event class selected scaling of flow coefficients in rapidity}
With this splitting in different event classes in hand we investigate now the multiplicative relations between the lower flow harmonics with the higher order harmonics that were previously found, namely $v_3\approx v_1\cdot v_2$ and $v_4 \approx 0.5 v_2^2$ (we refer to \cite{Nara:2018ijw} for a deeper discussion of the $v_4$ scaling relation). Especially the scaling of $v_3$ is interesting, because at ultra-relativistic collision energies the triangular flow has been shown to be mainly sensitive to the initial state \cite{ALICE:2011ab,Alver:2010gr,Petersen:2010cw}, while at the low energies investigated here where flow is extracted with respect to the first order event plane, triangular flow is mostly attributed to geometry and the intricate space and time dependent emission pattern of hadrons \cite{Hillmann:2018nmd,Mohs:2020awg}. 

Fig. \ref{fig:scaling_relations} shows the rapidity dependence of triangular flow $v_3$ (upper left) in comparison to the multiplicative relation $v_1\cdot v_2$ (upper right) and the rapidity dependence of quadrangular flow $v_4$ (lower left) in comparison to the scaling relation $0.5v_2^2$ (lower right) in the different event classes (from the most positive $v_2$ class in red to most negative $v_2$ class in purple, see legend). The integrated flow coefficient, i.e. without event class selection is shown as a black line. All calculations are done for 20-30\% peripheral Au+Au collisions with hard EoS at 1.23~AGeV. For the triangular flow, a nearly perfect matching with the product of directed and elliptic flow in all event classes can be observed. A similar observation can be made for the scaling of the quadrangular flow with the square of the elliptic flow. One should note that this scaling is remarkable because the elliptic flow changes sign as a function of event class, but still its square stays proportional to the $v_4$ flow harmonic and even the change of shape in rapidity from convex to concave is reproduced. Nevertheless a systematic numerical scaling factor is necessary which can be quantified as $v_4/v_2^2\approx 0.5$. 

How can we interpret these results? First of all, the results demonstrate that higher flow harmonics ($n>2$) are strongly intertwined with the first and second flow harmonics emphasizing the influence of geometry. In addition the quadrangular flow and its scaling with the elliptic flow was suggested to provide specific information on the applicability of ideal fluid dynamics \cite{Borghini:2005kd}. In \cite{Borghini:2005kd} the authors argued, using a hydrodynamic model, that a scaling relation of $v_4=0.5v_2^2$, as found in the present studies (see also \cite{Hillmann:2019wlt,Nara:2018ijw}), suggests applicability of ideal hydrodynamics, which means local equilibrium seems to be achieved. It should however be noted that such a conclusion is at variance with a study of the viscosity at HADES energies suggesting a substantial time dependent viscosity over entropy ratio \cite{Reichert:2020oes} and also not supported by the analysis of Ref. \cite{Nara:2018ijw}. 

\subsection{Event class selected analysis of flow shapes at midrapidity}
\begin{figure}
    \centering
    \includegraphics[width=\columnwidth]{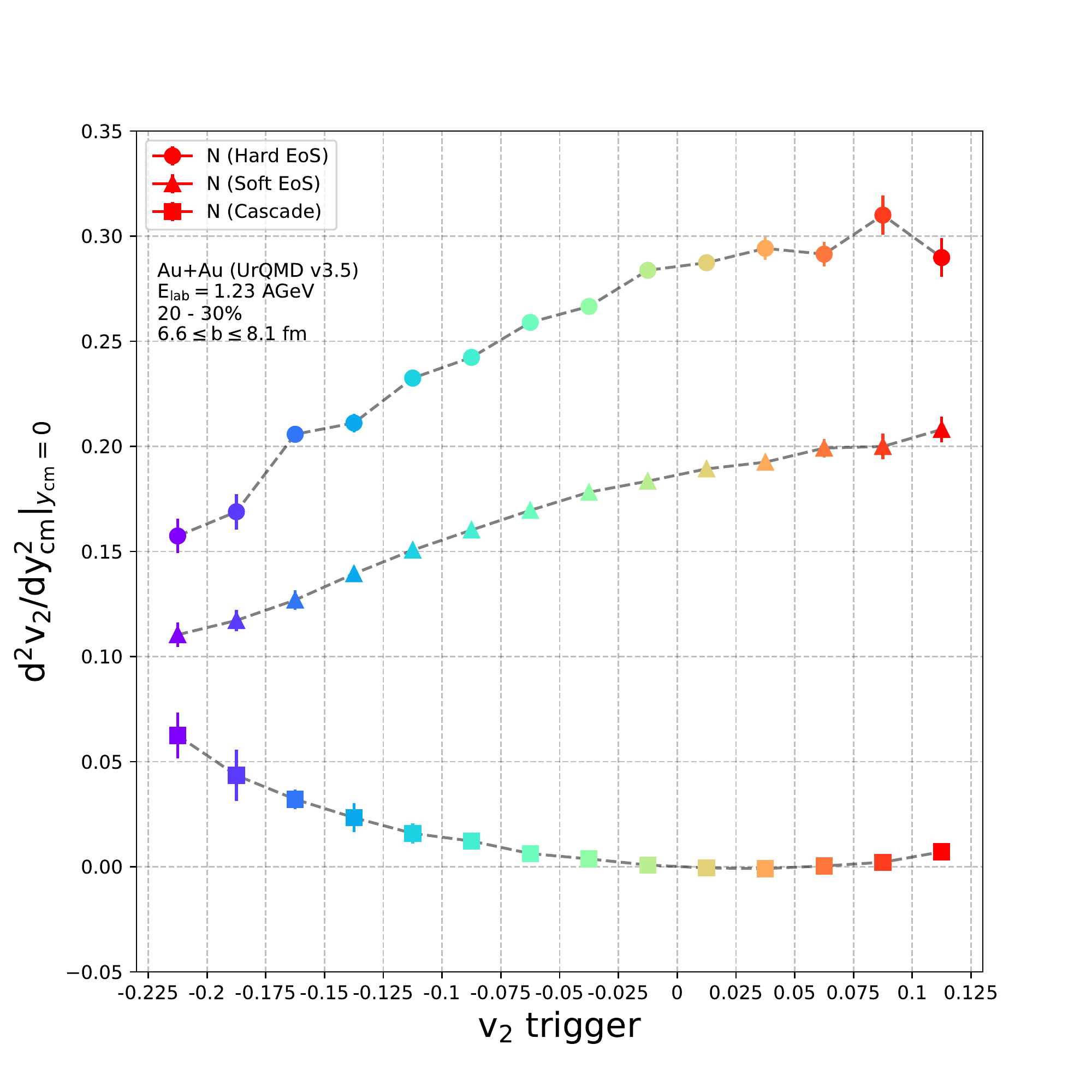}
    \caption{[Color online] The curvature of the elliptic flow component $v_2$ of nucleons at $y_\mathrm{cm}=0$ as a function of the final state elliptic flow $v_2$ (colored full symbols) integrated over $-0.5\leq y_\mathrm{cm}\leq0.5$ for a hard EoS (circles), a soft EoS (triangles) and in cascade mode (squares) from 20-30\% peripheral Au+Au collisions at 1.23~AGeV from UrQMD. }
    \label{fig:v2_slope}
\end{figure}

Let us finally use our event class analysis to obtain more information on the Equation-of-State. We start from the correlation of the $v_2$ shape in rapidity in dependence of the $v_2$ trigger. Specifically we propose to use the curvature of the elliptic flow around midrapidity, i.e. $\mathrm{d}^2v_{2}/\mathrm{d}y_\mathrm{cm}^2|_{y_\mathrm{cm}=0}$. Numerically, the curvature is obtained by fitting the rapidity dependence of $v_2$ with a quadratic polynomial and extracting the curvature from the fit. Fig. \ref{fig:v2_slope} shows the curvature of the elliptic flow component $v_2$ of nucleons at $y_\mathrm{cm}=0$ as a function of the final state elliptic flow $v_2$ (colored full symbols) integrated over $-0.5\leq y_\mathrm{cm}\leq0.5$ for a hard EoS (circles), a soft EoS (triangles) and in cascade mode (squares) from 20-30\% peripheral Au+Au collisions at 1.23~AGeV from UrQMD. The curvature of the elliptic flow is increasing with increasing trigger $v_2$ for the hard and soft equations-of-state while the curvature in case of the cascade simulation is decreasing with increasing $v_2$ event class.  This suggests that the curvature of the elliptic flow is strongly sensitive to the nuclear EoS.
\begin{figure}
    \centering
    \includegraphics[width=\columnwidth]{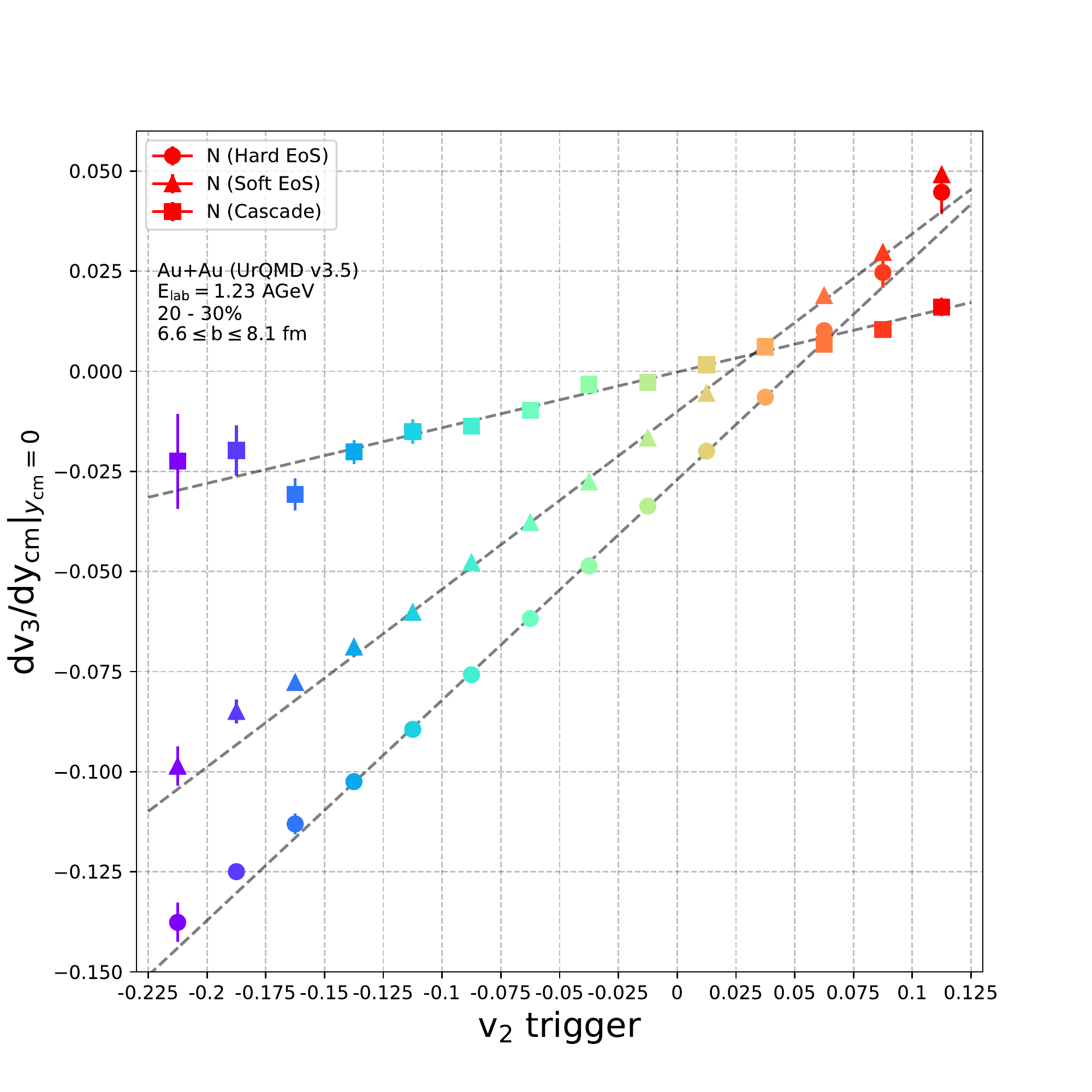}
    \caption{[Color online] The slope of the triangular flow component $v_3$ of nucleons at $y_\mathrm{cm}=0$ as a function of the final state elliptic flow $v_2$ (colored full symbols) integrated over $-0.5\leq y_\mathrm{cm}\leq0.5$ for a hard EoS (circles), a soft EoS (triangles) and in cascade mode (squares) from 20-30\% peripheral Au+Au collisions at 1.23~AGeV from UrQMD. }
    \label{fig:v3_slope}
\end{figure}

Next, we turn to the triangular flow shape and its correlation with the $v_2$ trigger. We specifically propose to use the correlation between the final $v_2$ and the slope $\mathrm{d}v_3/\mathrm{d}y_\mathrm{cm}|_{y_\mathrm{cm}=0}$ at midrapidity for the different event classes. Given our scaling assumption one would assume in first approximation that $\mathrm{d}v_3/\mathrm{d}y_\mathrm{cm}|_{y_\mathrm{cm}=0}= \mathrm{d}v_1/\mathrm{d}y_\mathrm{cm}|_{y_\mathrm{cm}=0} \cdot v_2$. Thus we expect to observe a linear dependence of $\mathrm{d}v_3/\mathrm{d}y_\mathrm{cm}|_{y_\mathrm{cm}=0}$ on $v_2$. In Fig. \ref{fig:v3_slope} we show $\mathrm{d}v_3/\mathrm{d}y_\mathrm{cm}|_{y_\mathrm{cm}=0}$ as a function of the $v_2$ event class for the nucleons in semi-peripheral Au+Au collisions at 1.23~AGeV for cascade calculations (no potential, squares), a soft EoS (triangles) and a hard EoS (circles) as full colored symbols. For all studied equations-of-state a linear dependence between $v_2$ and $\mathrm{d}v_3/\mathrm{d}y_\mathrm{cm}|_{y_\mathrm{cm}=0}$ is observed. The slope of the correlation depends on the stiffness of the EoS, a stiffer EoS shows a stronger incline. This correlation observable allows therefore also to pin down the equation-of-state more precisely than is usually possible.

\begin{figure}
    \centering
    \includegraphics[width=\columnwidth]{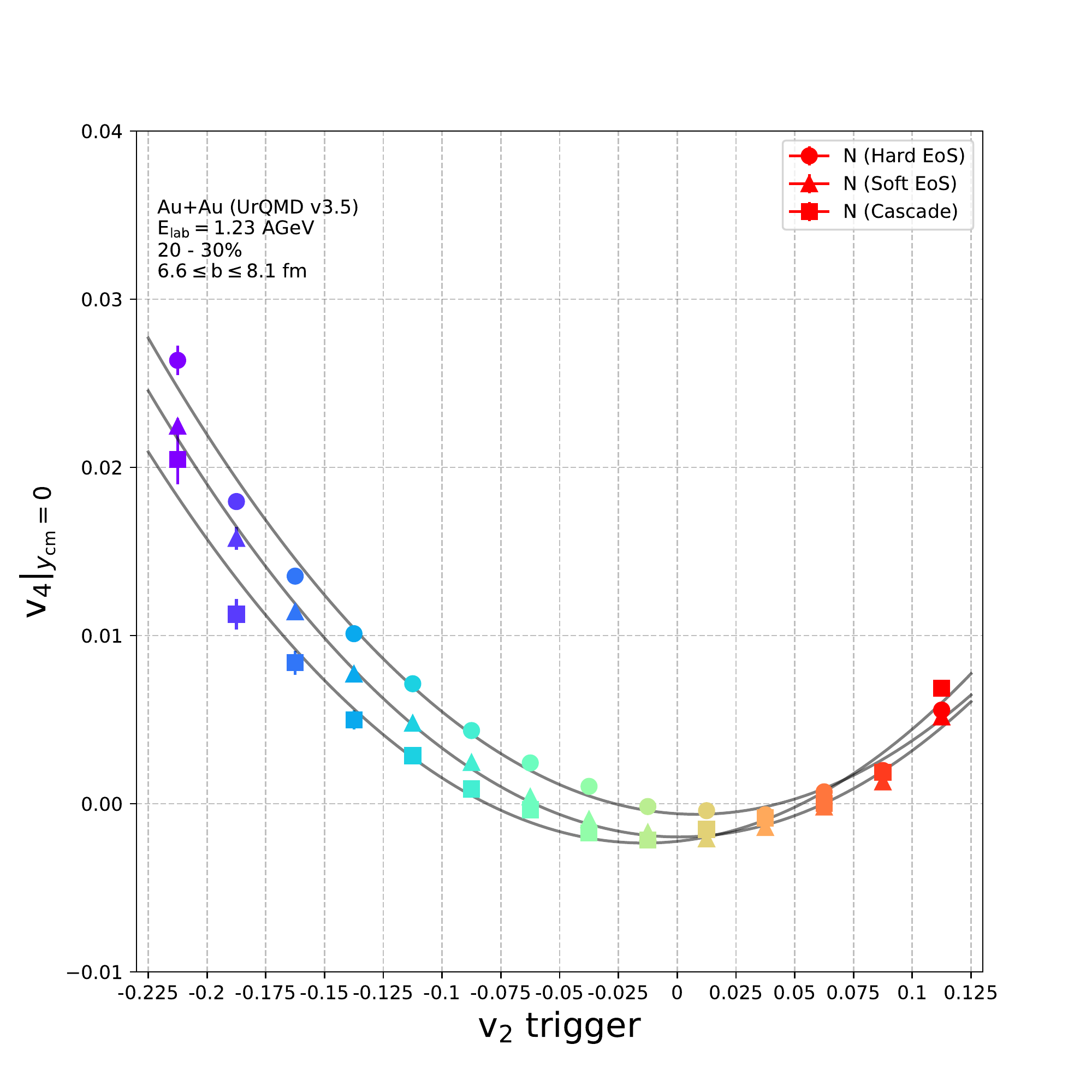}
    \caption{[Color online] The quadrangular flow component $v_4$ of nucleons at $y_\mathrm{cm}=0$ as a function of the final state elliptic flow $v_2$ (colored full symbols) integrated over $-0.5\leq y_\mathrm{cm}\leq0.5$ for a hard EoS (circles), a soft EoS (triangles) and in cascade mode (squares) from 20-30\% peripheral Au+Au collisions at 1.23~AGeV from UrQMD. }
    \label{fig:v4_v2}
\end{figure}

Last, we turn to the quadrangular flow. Motivated by measurements of the ATLAS collaboration \cite{ATLAS:2015qwl}, we analyze the quadrangular flow at midrapidity as a function of the elliptic flow trigger. Fig. \ref{fig:v4_v2} shows the quadrangular flow component $v_4$ of nucleons at $y_\mathrm{cm}=0$ as a function of the final state elliptic flow $v_2$ (colored full symbols) integrated over $-0.5\leq y_\mathrm{cm}\leq0.5$ for a hard EoS (circles), a soft EoS (triangles) and in cascade mode (squares) from 20-30\% peripheral Au+Au collisions at 1.23~AGeV from UrQMD. The lines depict least-squares quadratic fits. We observe a quadratic dependence of $v_4$ on the $v_2$ trigger as seen by the quadratic polynomial fit functions. The coefficients of the quadratic terms are 0.521 (hard EoS), 0.521 (soft EoS) and 0.522 (cascade) which again confirm the scaling relation $v_4\propto 0.5v_2^2$. However, the value of $v_4$ alone is not sensitive to the EoS.

\begin{figure}
    \centering
    \includegraphics[width=\columnwidth]{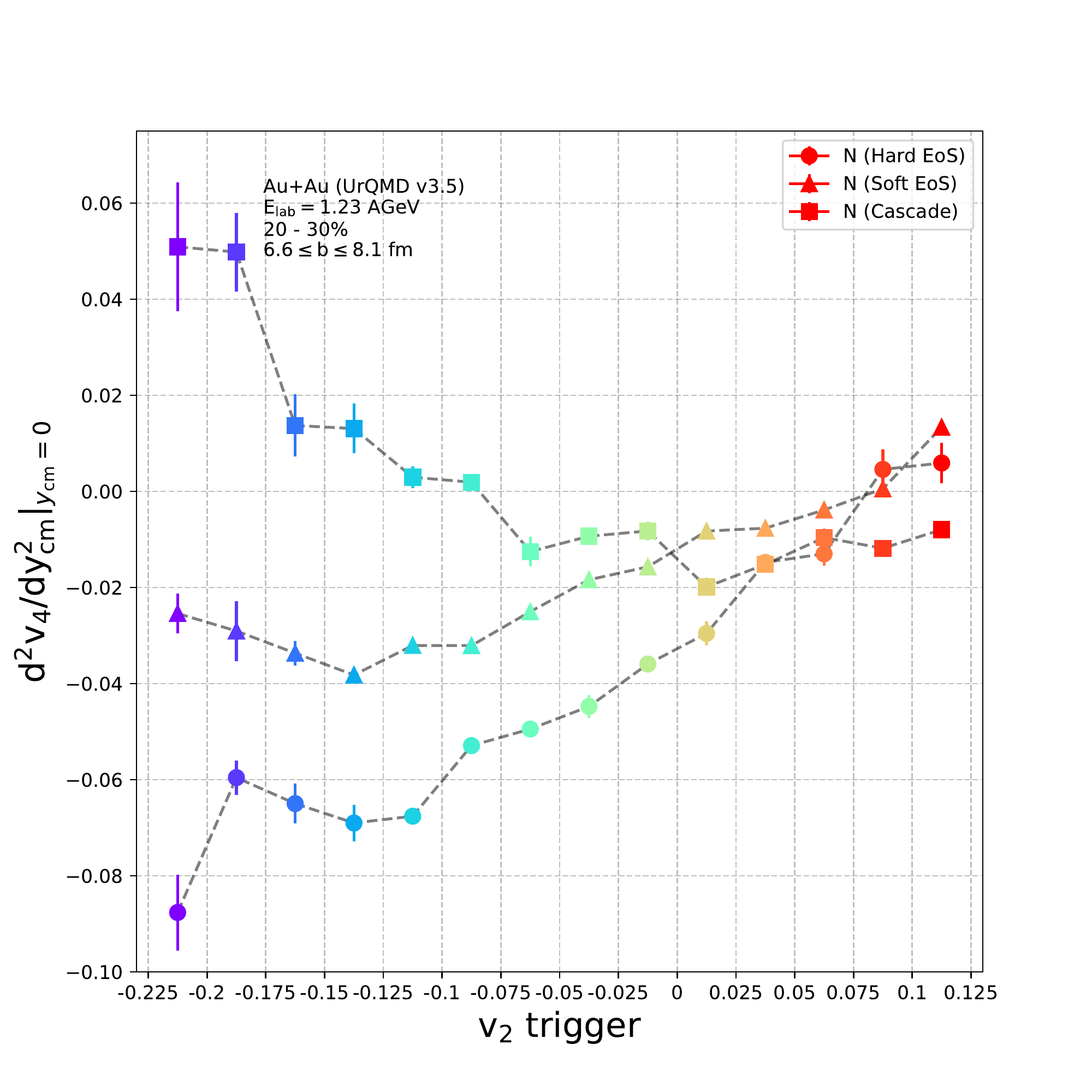}
    \caption{[Color online] The curvature of the quadrangular flow component $v_4$ of nucleons at $y_\mathrm{cm}=0$ as a function of the final state elliptic flow $v_2$ integrated over $-0.5\leq y_\mathrm{cm}\leq0.5$ for a hard EoS (circles), a soft EoS (triangles) and in cascade mode (squares) from 20-30\% peripheral Au+Au collisions at 1.23~AGeV from UrQMD. }
    \label{fig:v4_slope}
\end{figure}

Therefore, we perform a similar analysis as for the elliptic flow and extract the $v_4$ shape in rapidity as a function of the $v_2$ trigger. Again, we use a quadratic polynomial and extract the curvature from the fit, i.e. $\mathrm{d}^2v_{4}/\mathrm{d}y^2_\mathrm{cm}|_{y_\mathrm{cm}=0}$. Fig. \ref{fig:v4_slope} shows the curvature of the quadrangular flow  of nucleons at $y_\mathrm{cm}=0$ as a function of the final state elliptic flow $v_2$ (colored full symbols) integrated over $-0.5\leq y_\mathrm{cm}\leq0.5$ for a hard EoS (circles), a soft EoS (triangles) and in cascade mode (squares) from 20-30\% peripheral Au+Au collisions at 1.23~AGeV from UrQMD. We observe a strong splitting of the quadrangular flow shapes towards negative trigger $v_2$, which indicates a strong dependence on the nuclear equation of state.

\section{Flow correlation functions}
\begin{figure}
    \centering
    \includegraphics[width=\columnwidth]{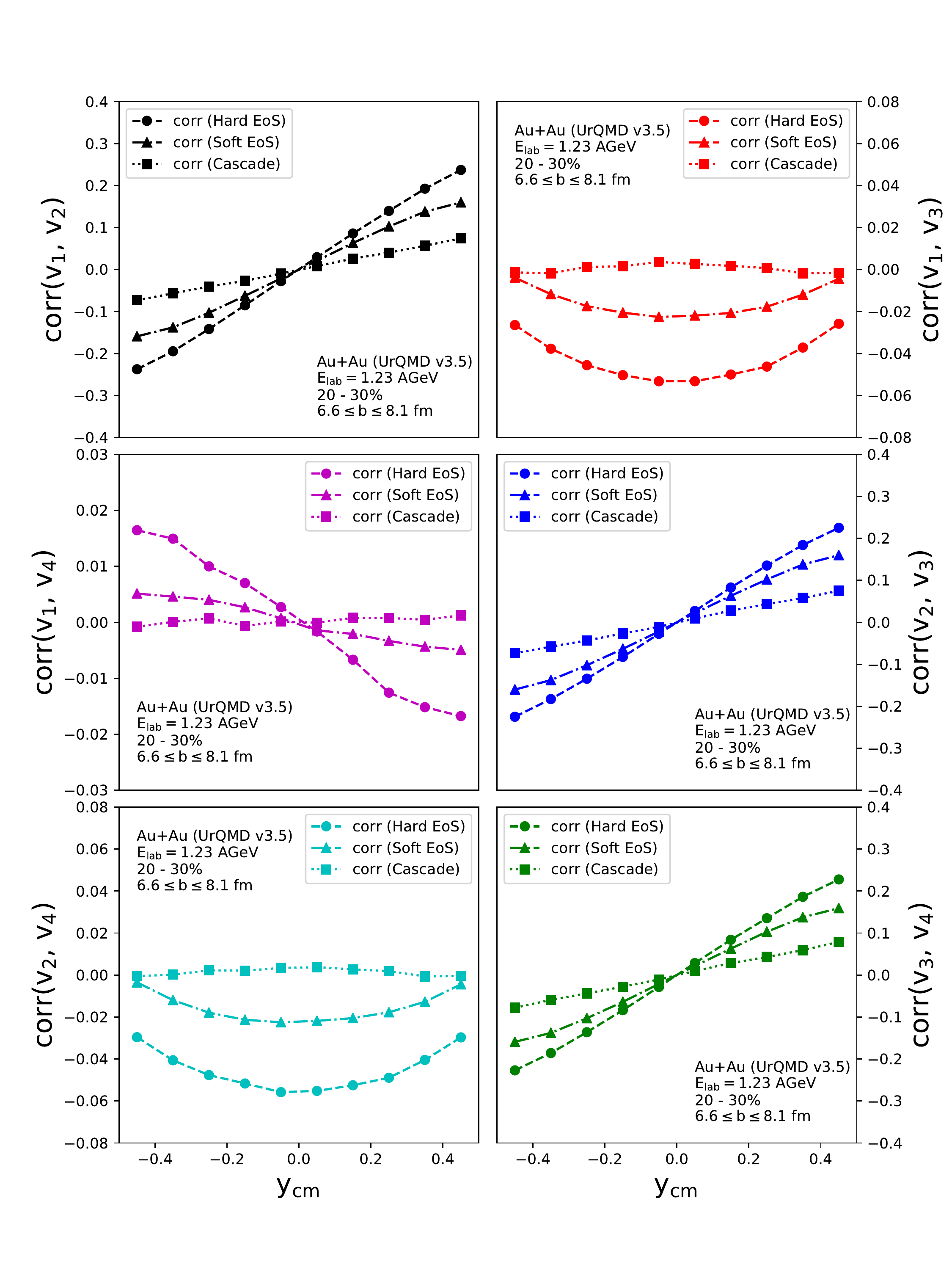}
    \caption{[Color online] The Pearson correlation function $corr(v_n,v_m)$ (full symbols) between the first four flow harmonics of nucleons as a function of rapidity for a hard EoS (circles), a soft EoS (triangles) and in cascade mode (squares) from 20-30\% peripheral Au+Au collisions at 1.23~AGeV from UrQMD.}
    \label{fig:flow_corr}
\end{figure}

To bridge the gap to previous studies of flow fluctuations and their correlations \cite{Niemi:2012aj,Zhou:2015eya,Zhu:2016puf} we finally investigate the correlation functions among the flow harmonics. To this aim, we use the linear correlation function $corr(v_n,v_n)$ (also known as the Pearson coefficient\footnote{Let us note, that the Pearson coefficient provides a measure for linear dependence of two random variables ($|corr|=1$ implies perfect linear dependence), and that a vanishing Pearson coefficient does not rule out any nonlinear correlation.}) between the first four flow harmonics calculated as
\begin{equation}
    corr(v_n,n_m) = \frac{\langle v_nv_m\rangle - \langle v_n\rangle\langle v_m\rangle}{\sigma_{v_n}\sigma_{v_m}}.
\end{equation}
Here, the standard deviation $\sigma_{v_i}=\sqrt{\langle v_i^2\rangle-\langle v_i\rangle^2}$ is used to normalize the covariance. 
In Fig. \ref{fig:flow_corr} we show the Pearson correlation function $corr(v_n,v_m)$ (full symbols) between the first four flow harmonics of nucleons as a function of rapidity for a hard EoS (circles), a soft EoS (triangles) and in cascade mode (squares) from 20-30\% peripheral Au+Au collisions at 1.23~AGeV from UrQMD.

We observe a strongly pronounced rapidity dependent correlation between the first and second, the second and third, and the third and forth flow harmonic which is point-symmetric around $y_\mathrm{cm}=0$, while the correlation between the first and third, and second and fourth flow harmonic is symmetric around $y_\mathrm{cm}=0$ but its value is negative and smaller, even vanishing for the cascade simulation. The correlation of the first and fourth flow harmonic is negligibly small. Especially, the previously observed scaling relation of quadrangular flow ($v_4\propto v_2^2$) is interesting to discuss. The correlation is clearly observed in our calculations and has also been extensively studied in various publications \cite{Nara:2018ijw,Borghini:2005kd,Hillmann:2019wlt}, but due to its quadratic dependence the Pearson correlation shows no significant signal, but only a maximal value of -0.06. Although the Pearson correlation allows at most to draw conclusions about linear dependence, the results demonstrate nonetheless a dependence of the Pearson coefficient on the employed Equation-of-State: the stiffer the EoS the stronger the correlation becomes for all harmonic flow combinations. Together with the slope of triangular flow and the curvature of elliptic and quadrangular flow shown above this composes a neat possibility to pin down the nuclear Equation-of-State with high precision. 

\section{Conclusion}
We have employed the Ultra-relativistic Quantum-Molecular-Dynamics model (UrQMD v3.5) to study semi-peripheral Au+Au collisions (20-30\% centrality) at a beam energy of 1.23~AGeV. We found that the final state $v_2$ of a single event fluctuates strongly around its mean value allowing to identify classes of events with selected elliptic flow. We employ these event classes as a trigger to extract the rapidity dependence of the first to fourth order harmonic flow coefficients. Using this trigger, the directed flow is only slightly affected in its magnitude, while the elliptic flow acquires (as expected) a linear shift with positive correlation. The higher order flow components however reveal interesting new features, namely as a function of rapidity the triangular flow changes sign and the quadrangular flow changes shape from concave to convex when going from negative to positive $v_2$ event classes. All event classes are found to fulfill the scaling relation $v_3\propto v_1v_2$ nearly perfectly, also the quadrangular flow exhibits the ideal fluid scaling $v_4 = 0.5 v_2^2$ around midrapidity in all event classes. We further demonstrated that, correlation between the slope of the triangular flow and the curvature of the elliptic and quadrangular flow at midrapidity as a function of the selected elliptic flow are highly sensitive to the EoS. We thus propose these features as novel tools to obtain further information on the nuclear Equation-of-State. Lastly, we investigated the Pearson coefficients which underline the correlations analyzed with the event class selection and allow to compare the current analysis to calculations at higher collision energies.

\begin{acknowledgements}
The authors thank Paula Hillmann and Behruz Kardan for fruitful discussion about the flow harmonics and the analysis. This article is part of a project that has received funding from the European Union’s Horizon 2020 research and innovation programme under grant agreement STRONG – 2020 - No 824093. J.S. thanks the Samson AG for funding. Computational resources were provided by the Center for Scientific Computing (CSC) of the Goethe University and the ``Green Cube" at GSI, Darmstadt. This project was supported by the DAAD (PPP Thailand). This research has received funding support from the NSRF via the Program Management Unit for Human Resources \& Institutional Development, Research and Innovation [grant number B16F640076].
\end{acknowledgements}



\end{document}